\documentclass[10pt,conference]{IEEEtran}
\usepackage{graphicx}

\begin{document}

\title{Matrix Construction Using Cyclic Shifts of a Column}
\author{\authorblockN{Andrew Z Tirkel}
\authorblockA{School of Mathematical Sciences\\
Monash University\\
PO Box 28M, Victoria 3800, Australia \\
Email: atirkel@bigpond.net.au} \and
\authorblockN{Tom E Hall}
\authorblockA{School of Mathematical Sciences\\
Monash University\\
PO Box 28M, Victoria 3800, Australia\\
Email: tom.hall@sci.monash.edu.au}}
\maketitle

\begin{abstract}
This paper describes the synthesis of matrices with good
correlation, from cyclic shifts of pseudonoise columns. Optimum
matrices result whenever the shift sequence satisfies the constant
difference property. Known shift sequences with the constant (or
almost constant) difference property are: Quadratic (Polynomial) and
Reciprocal Shift modulo prime, Exponential Shift, Legendre Shift,
Zech Logarithm Shift, and the shift sequences of some m-arrays. We
use these shift sequences to produce arrays for watermarking of
digital images. Matrices can also be unfolded into long sequences by
diagonal unfolding (with no deterioration in correlation) or
row-by-row unfolding, with some degradation in correlation.
\end{abstract}

\section{Introduction}
Sequences with good auto and cross-correlation have many
applications in modern communications, whilst radar, sonar and
digital watermarking of images also requires arrays (matrices) with
similar properties. Some sequences can be naturally folded into
arrays, whilst arrays can always be unfolded into sequences. Here,
we consider the synthesis of arrays with good correlation
properties. We focus on arrays where each column is a cyclic shift
of one pseudonoise (or impulse) column, or a constant column. The
shift sequence of an array lists the cyclic shift of each
pseudonoise or impulse column, and lists - for each constant column.
A sequence is considered pseudonoise if its autocorrelation takes on
two values, a high value at zero shift and a single low value for
all other shifts. Such sequences have been studied extensively,
because of their applications to communications and radar. The
matrix auto and cross-correlations are summations of the auto and
cross-correlations of the overlaying columns. For matrices
constructed from the same column sequence, these are all
autocorrelations of that column and hence are $-1$ or $v$ ($v$ is
the column length) for pseudonoise columns and $1$ or $0$ for
impulse columns. The auto and cross-correlations are then determined
by the numbers of matching columns. Shift sequences which result in
0 or 1 matching column are particularly prized. A two-dimensional
cyclic shift of a matrix by $(k,l)$ is obtained by rotating the rows
by $k$ places to the right and each column by $l$ places downwards.
The correlation of two matrices is the list of the inner products of
one matrix with all the two-dimensional cyclic shifts of the other
matrix.  In our context, the correlation of two matrices is
calculated by counting the number of matching columns for each
$(k,l)$ shift. The paper presents an analysis of matrices
constructed by using shift sequences based on: polynomials,
exponentials, logarithms and the naturally occurring shift sequences
of m/GMW arrays.
\section{Polynomial Shift Sequences}
Consider a shift sequence calculated as a polynomial of order $n>1$,
($\varphi _{1}(x)$ modulo $p$), with coefficients from $Z_p $:
\begin{equation}\label{1}
\varphi _1 (x)=a_n x^n+a_{n-1} x^{n-1}+...+a_1 x^1+a_0
\end{equation}
When column labeled \textit{x=0,1,{\ldots}, p-1}, is cyclically
shifted through $\varphi_{1} (x)$ places downwards, this describes a
$p\times p$ matrix ${\textbf{\emph{A}}_{\textbf{\emph{1}}}}$. There
are $p^{2}-1$ other matrices cyclically equivalent to
${\textbf{\emph{A}}_{\textbf{\emph{1}}}}$. These are obtained by all
two-dimensional non-zero shifts to matrix
${\textbf{\emph{A}}_{\textbf{\emph{1}}}}$. Horizontal shifts are
equivalent to the transformation $x'= x+k$, whilst vertical shifts
are achieved by the transformation $\varphi '(x)=\varphi (x)+l$
where $k$ and $l$ are chosen from $Z_{p}$. Shift polynomials which
differ only in $a_{0}$ and/or $a_{n-1}$ describe matrices, which are
cyclic shifts of each other. There are $p^{n-1}-1$ inequivalent
matrices.
\subsection{Autocorrelation}
The autocorrelation for shift $(k,l)$ is obtained from the number of
matching columns: i.e. solutions of
\begin{equation}\label{2}
    \varphi _1 (x)-\varphi' _1 (x')=0
\end{equation}
The above is a polynomial of degree at most $n-1$ and this is an
upper bound on the number of matching columns. For pseudonoise
columns with peak autocorrelation of $p$ and off-peak
autocorrelation of $-1$, the off-peak matrix autocorrelation takes
on values restricted to: $-p, +1, p+2, 2p+3,...,(n-2)p+(n-1)$. The
frequency of occurrence of each autocorrelation value depends on the
number of distinct roots of the difference polynomial for the shift
$(k,l)$. The lowest meaningful value of $n=2$ [1] results in
three-valued matrix autocorrelation: $p^{2},-p,+1.$ In case of a
ternary Legendre column, with autocorrelation values of $p-1$ and
$-1$, the matrix autocorrelation values become: $p(p-1),0,-p$. The
quadratic shift sequence is one which possesses the distinct
difference property (DDP), in that all differences in the shift
sequence taken a fixed distance apart appear exactly once. This is a
result of the (linear) difference polynomial having exactly one
solution for each $(k,l)$ shift. Subsequent sections describe other
known shift sequences with DDP, where each difference is allowed to
appear at most once.
\subsection{Cross-correlation}
Consider another matrix, ${\textbf{\emph{A}}_{\textbf{\emph{2}}}}$
built from identical column sequences, using the shift sequence:
\begin{equation}\label{3}
    \varphi _2 (x)=b_m x^m+b_{m-1} x^{m-1}+...+b_1 x^1+b_0
\end{equation}
The number of matching columns between the matrices is obtained by
solving
\begin{equation}\label{4}
\varphi _2 (x)-\varphi' _1 (x')=0
\end{equation}
where $x'$ and $\varphi'$ are defined as above.  The difference
polynomial is of degree at most \textit{max(m,n)}, so this is an
upper bound on the number of matching columns.\\
\emph{Example 1}\\
 A quadratic shift sequence is:
\textit{0,6,4,1,4,6,0 mod 7}. A matrix built from this shift
sequence using a ternary Legendre sequence column is shown below.
\begin{table}[htbp]
\begin{center}
\begin{tabular}{|l|l|l|l|l|l|l|}
\hline {\small 0}& {\small 1}& {\small -1}& {\small -1}& {\small
-1}& {\small 1}&
{\small 0} \\
\hline {\small 1}& {\small 1}& {\small 1}& {\small 0}& {\small 1}&
{\small 1}&
{\small 1} \\
\hline {\small 1}& {\small -1}& {\small -1}& {\small 1}& {\small
-1}& {\small -1}&
{\small 1} \\
\hline {\small -1}& {\small 1}& {\small -1}& {\small 1}& {\small
-1}& {\small 1}&
{\small -1} \\
\hline {\small 1}& {\small -1}& {\small 0}& {\small -1}& {\small 0}&
{\small -1}&
{\small 1} \\
\hline {\small -1}& {\small -1}& {\small 1}& {\small 1}& {\small 1}&
{\small -1}&
{\small -1} \\
\hline {\small -1}& {\small 0}& {\small 1}& {\small -1}& {\small 1}&
{\small 0}&
{\small -1} \\
\hline
\end{tabular}
\label{tab1}
\end{center}
\end{table}
The autocorrelation off-peak values are $0,-p$. Another matrix can
be constructed from the quadratic shift sequence:
\textit{0,5,1,2,1,5,0. }It is shown below, as well as its
cross-correlation with the first matrix, which is constrained to
\textit{0,-(p-1),+(p-1)}.
\begin{table}[htbp]
\begin{center}
\begin{tabular}{|l|l|l|l|l|l|l|}
\hline {\small 0}& {\small 1}& {\small -1}& {\small -1}& {\small
-1}& {\small 1}&
{\small 0} \\
\hline {\small 1}& {\small -1}& {\small 0}& {\small -1}& {\small 0}&
{\small -1}&
{\small 1} \\
\hline {\small 1}& {\small 1}& {\small 1}& {\small 0}& {\small 1}&
{\small 1}&
{\small 1} \\
\hline {\small -1}& {\small -1}& {\small 1}& {\small 1}& {\small 1}&
{\small -1}&
{\small -1} \\
\hline {\small 1}& {\small -1}& {\small -1}& {\small 1}& {\small
-1}& {\small -1}&
{\small 1} \\
\hline {\small -1}& {\small 0}& {\small 1}& {\small -1}& {\small 1}&
{\small 0}&
{\small -1} \\
\hline {\small -1}& {\small 1}& {\small -1}& {\small 1}& {\small
-1}& {\small 1}&
{\small -1} \\
\hline
\end{tabular}
\label{tab2}
\end{center}
\end{table}
There are $p-1$ cyclically distinct quadratic matrices, with the
same bounds on all autocorrelations and cross-correlations. This
construction is preferred where large numbers of matrices, with
minimum interference are required. A disadvantage is low
cryptographic security, owing to low linear complexity of the shift
sequence. Quadratic shift sequence matrices can be punctured,
columns inserted columns shortened or lengthened, with predictable
results [4].
\subsection {Column Sequences}
The required sequence length of $p$ can be satisfied by a Legendre
sequence, Hall sequence or binary m-sequence.
\section{Exponential Shift Sequence}
An exponential shift sequence, with $p-1$ entries of the form
$g^{i}$, where $g$ is any primitive root in $Z_p$, has the distinct
difference property [2]. There are $\phi (p-1)$ sequences, where
$\phi$ is the Euler Totient Function. The treatment in \S\S  II,
III, of exponential and logarithmic functions, as shift sequences,
is an adaptation of that for Costas Arrays [3]. Consider two arrays
constructed using common column sequences and the shift sequences:
$\varphi _1 (j)=g^j$ and $\varphi _2 (j)=h^j$, where g and h are
primitive roots of $Z_p $. Columns match whenever
 \begin{equation}\label{5}
    h^j=g^{j+k}+l
\end{equation} But
$h=g^r$ where $r\in Z_p $ and \textit{gcd(r,p-1)=1.}The columns
which match are solutions of
\begin{equation}\label{6}
    g^{rj}-g^{j+k}-l=0
\end{equation}
Let $X=g^j$. Therefore
 \begin{equation}\label{7}
    X^r-g^kX-l=0
\end{equation}
Equation (7) has at most $r$ solutions. Hence, an upper bound on the
number of matching columns is $r$.
\subsection{Autocorrelation}
Here $r=1$ so the number of matching columns is: $p-1,1,0$. The
matrix autocorrelation values are: $p^{2}-p,+2,1-p$.
\subsection{Cross-correlation}
The best cross-correlation is achieved for $r=-1$. Then, equation
(7) becomes:
\begin{equation}\label{8}
    X^{-1}-g^kX-l=0 \quad or \quad  g^{k}X^{2}+X-1=0
\end{equation}
Hence, an upper bound on the number of matching columns is $2$. The
matrix cross-correlation values are: $p+3,+2,1-p$.
\subsection{Column Sequences}
The required sequence of length $p$ can be any Legendre sequence,
Hall sequence or some binary m-sequences.
\section{Logarithmic shift}
Two types of discrete logarithms arise:
 \begin{enumerate}
    \item index function giving rise to Legendre Shift Arrays
    \item Zech logarithm giving rise to Zech Shift Arrays
\end{enumerate}
The former applies to $Z_{p}$, whilst the latter to $GF(p^{m})$
\subsection{Legendre Arrays}
Let a pseudonoise sequence of length $(p-1)$ be shifted cyclically
as a function of column number $j$:
\begin{equation}\label{9}
    \varphi_{r} (j)=r\times ind_{g}(j), \textit{where}\quad  r\in
    1,2,...,p-1
\end{equation}
 with $g$
being a primitive root of $Z_p$ and $ind_{g}$ being the $k 0\leq
k\leq p-2$, such that $g^{k}=j$. Note that $\varphi (j)$ is
expressed \textit{mod (p-1)}. The column indices ($j)$ of the array
range from $0$ to $p-1$, so the matrix has $p$ columns with $(p-1)$
rows, with the first column being blank, since $ind(0)$ is not
defined. There are $p-2$ sequences. For example, for $p=7, g=3$, the
shift sequence is: \textit{-,0,2,1,4,5,3.}
\subsection{Correlation}
An appropriate choice of $g$ can make $r=1$ for one of the shift
sequences. For two shift sequences $\varphi_{r} (j)$ and
$\varphi_{1} (j)$ the number of matching columns is the number of
solutions of:
\begin{equation}\label{10}
r\times ind_{g}(j)- ind_{g}(j+k)-l=0
\end{equation}
or
\begin{equation}\label{11}
 j^r-g^lj+g^lk=0 \end{equation}
\subsubsection {Autocorrelation}
For autocorrelation, $r=1$ and so, the number of matching columns is
$p-1$ for $k=l=0$, $1$ for $k \ne 0$ or $0$ for $k=0, l \ne 0$ and
$l=0,k \ne 0$. The matrix autocorrelation values are: $(p-1)^{2})$
for full match, $+1$ for $1$ column match, $-(p-1)$ for $0$ column
match. These values can be altered, if the blank column is
substituted by a constant column.
\subsubsection{Cross-correlation}
The best result of an upper bound of $2$ columns matching is
obtained for $r=2$ or $-1$. This occurs for pairs of sequences only.
The matrix cross-correlation values are: $p+1,+1,-(p-1)$.
\subsubsection{Column Sequence}
A suitable column sequence for such a construction is any m-sequence
over $Z_p$ and its mapping onto roots of unity $S(i)=\omega ^{g^i}$,
Schroeder [2,26.19], where $g$ is a primitive root of $Z_p$. $\omega
=e^{\frac{2\pi i}{p}}$ is a primitive root of unity. Here $p=7.$
Multiples of this angle are shown.
\[
\left[ {{\begin{array}{*{20}c}
 - \hfill & 5 \hfill & 3 \hfill & 1 \hfill & 6 \hfill & 4 \hfill & 2 \hfill
\\
 - \hfill & 4 \hfill & 1 \hfill & 5 \hfill & 2 \hfill & 6 \hfill & 3 \hfill
\\
 - \hfill & 6 \hfill & 5 \hfill & 4 \hfill & 3 \hfill & 2 \hfill & 1 \hfill
\\
 - \hfill & 2 \hfill & 4 \hfill & 6 \hfill & 1 \hfill & 3 \hfill & 5 \hfill
\\
 - \hfill & 3 \hfill & 6 \hfill & 2 \hfill & 5 \hfill & 1 \hfill & 4 \hfill
\\
 - \hfill & 1 \hfill & 2 \hfill & 3 \hfill & 4 \hfill & 5 \hfill & 6 \hfill
\\
\end{array} }} \right]
\]
\subsection{Zech Arrays}
Logarithmic arrays of size $(p^m-1)\times (p^m-1)$ can be
constructed using column shifts determined by the Zech logarithm of
elements of a Galois Field \textit{GF(p}$^{m})$. For the general
(Golomb) construction, the shift sequence is
\begin{equation}\label{12}
    \varphi (j)=\log _\beta (1-\alpha ^j)
\end{equation}
$\alpha$ and $\beta$ are primitive elements of
\textit{GF(p}$^{m})$.\\ The total number of distinct arrays is:
$\frac{[\phi (p^m-1)]^2}{m}$. The special case of $\beta =\alpha$ is
the Lempel construction.
\subsubsection{Correlation}
Consider two shift sequences:
\begin{equation}\label{13}
    \varphi _1 (j)=\log _\beta (1-\alpha ^j) \quad and \quad \varphi _2 (j)=\log
_\gamma (1-\delta ^j)
\end{equation}
where $\gamma =\beta ^r$and $\delta =\alpha ^s$\\
and \textit{gcd(r,p}$^{m}-1)$=\textit{ gcd(s,p}$^{m}-1)=1.$ We can
interpret $r$ as a multiplier and $s$ as a sampler or decimation. An
upper bound on the number of matching columns is determined by the
number of solutions of: \begin{equation}\label{14}   \log _\beta
(1-\alpha ^{j+k})+l=\log _\gamma (1-\delta^j)=r^{-1}[\log _\beta
(1-\alpha ^{sj})]\end{equation}
\begin{center}so $\log _\beta \frac{1-\alpha ^{j+k}}{(1-\alpha ^{sj})^{-r}}=-l$ and $1-\alpha
^{j+k}=\beta ^{-l}(1-\alpha ^{sj})^{-r}$\end{center} Finally
$(1-\alpha ^{j+k})^r=\beta ^{-l}(1-\alpha ^{sj})$
\\Once again, let $\alpha ^j=X$. Therefore
\begin{equation}\label{15}
    (1-\alpha ^kX)^r=\beta ^{-l}(1-X^s)
\end{equation}
(15) is a polynomial equation of degree \textit{max (r,s)}.
\subsubsection{Autocorrelation}
$r=s=1$, so the numbers of matching columns are: $p-1,1,0$.
\subsubsection{Cross-Correlation}
$r=-1, s=1$ or $r=1, s=-1$ or $r=2, s=1$ or $r=1, s=2$ once again
result in quadratic forms and hence limit the number of matching
columns to $0,1,2$.\\
\emph{Example 2} A shift sequence of length $15$ over \textit{GF
(2}$^{4})$ is:
\begin{center}-,4,8,14,1,10,13,9,2,7,5,12,11,6,3.\end{center}
Another shift sequence is:
\begin{center}-,8,1,13,2,5,11,3,4,14,10,9,7,12,6.\end{center} The lower sequence could be obtained by
decimation of the above sequence by $2$ or multiplication of the
entries by $2$. The cross-correlation shows that $2$ columns can
match at most.
\subsubsection{Column Sequences}
Any m-sequence/GMW sequence of length $p^{m}-1$ is suitable.
\section{Other Shift Sequences}
\subsection{Shift Sequences from M-arrays}
An m-sequence of length $p^{km}-1$ can be written row-by-row as a
matrix, such that all columns are constants or cyclic shifts of one
short m-sequence of length $p^m-1$. This matrix can be described by
a shift sequence and the column sequence. Formally, this shift
sequence $f_j$ can be obtained from finite field theory [10]:
\begin{equation}\label{16}
    f_j=ind_\gamma (Tr_{^m}^n (\alpha ^j)) \end{equation}
where $Tr^{n}_{m}$ is the trace function, which takes the sum of
conjugates (as defined in [11]). This shift sequence can be
converted into a perfect shift sequence for two-dimensional matrices
by the addition of an appropriate linear function [4]. Then, all
differences appear equally frequently for all cyclic shifts of the
shift sequence. Such shift sequences can be used to construct new
matrices, by substituting the columns by other pseudonoise
sequences. For example, an m-sequence of length $120$ ($p^2-1$, with
$p=11)$ can be written as an array of $10$ rows, each of length
$12$. Each column of the array is an m-sequence of length $10$,
except for the single null sequence. The twelve columns are
described by a shift sequence modulo $10$. This shift sequence can
be converted into a perfect shift sequence modulo $5$, [4]:
\textit{3,2,2,4,0,3,-,3,0,4,2,2. } This produces a perfect matrix,
by using it to shift (cyclically) a ternary Legendre sequence of
length 5: \textit{0,1,-1,-1,1}. The choice of $a=\sqrt
{\frac{11}{5}}$ for the entry $a$ in the constant column adjusts the
autocorrelation numbers to : $55$ for zero shift, and $0$ for all
other shifts, an example of perfect autocorrelation for the array
and its diagonal sequences.
\begin{table}[htbp]
\begin{center}
\begin{tabular}{|l|l|l|l|l|l|l|l|l|l|l|l|}
\hline -1& -1& -1& 1& 0& -1& $a$& -1& 0& 1& -1&
-1 \\
\hline -1& 1& 1& -1& 1& -1& $a$& -1& 1& -1& 1&
1 \\
\hline 1& 0& 0& -1& -1& 1& $a$& 1& -1& -1& 0&
0 \\
\hline 0& 1& 1& 1& -1& 0& $a$& 0& -1& 1& 1&
1 \\
\hline 1& -1& -1& 0& 1& 1& $a$& 1& 1& 0& -1&
-1 \\
\hline
\end{tabular}
\end{center}
\end{table}
\subsection{Hyperbolic Shift Sequences}
Hyperbolic sequences have been used for designing Frequency Hop
Patterns [8]. They can also be applied to construct matrices of the
type described in this paper.
\section{Window Property}
A column sequence is said to have the $n\times1$ strong window
property if each possible n'tuple occurs and occurs once only, as n
consecutive symbols in the column. The weak window property is that
no window of n symbols appears more than once. Each m-sequence as a
column sequence in this paper has the weak window property. The
shift sequences discussed in this paper also have the window
property: the quadratic shift sequence mod $p$ has the strong window
property, whilst all the others have the weak window property. For
each shift sequence in this paper, each doubleton $(\varphi(j),
\varphi(j+k))$ appears at most once. Hence, an array constructed
using such a shift sequence, with a column sequence having  the weak
$n\times1$ window property, has the weak $n\times2$ window property,
for any fixed separation of the two window columns [7]. Arrays with
window properties are useful as registration patterns in structured
light for medical imaging.
\section{Applications}
\subsection{Watermarking}
Matrices have been used as watermarks in various images in spatial
and transform domains [5]. Entries with real values are preferred,
although matrices over complex
numbers have also been embedded as watermarks [6].\\
To keep watermarks unobtrusive, and immune to correlation type
attack, the matrices need to be as large, and as efficient as
possible (least zero-value entries), with low off-peak
autocorrelation. This makes use of the maximum processing gain. All
matrices described in the previous sections satisfy these criteria.
Some applications require additional immunity to cryptographic
attack. Since most attacks involve linear processing, linear
complexity is a good measure. Matrices constructed using
exponential, logarithmic and hyperbolic shift sequences, applied to
high linear complexity column sequences e.g. GMW, Bent sequences etc
provide this feature. Other applications require large information
storage in the watermark. Typically, information is stored in the
value of the cyclic shift of the matrix. Information capacity can
only be increased by combining more matrices with low mutual
cross-correlation. The polynomial shift sequence appears to be the
only one capable of this. The example of this feature is illustrated
in the figures below. Figure 1 shows the result of adding 4
$127\times127$ matrices constructed using 4 different quadratic
shift sequences and the same m-sequence column. The matrices all
have different cyclic shifts, each one contributing a bit word to
the information content. Figure 2 shows the original image. Figure 3
shows the watermarked image, where the watermark is just visible.
Figure 4 shows the result of correlating the watermarked image with
the four matrices, showing 4 distinct peaks, one for each array.
Such a composite watermark can carry almost 4 ASCII characters. The
performance of such watermarks is much better for full scale images,
using RGB and can be extended to video.\\

\subsection{Communications}
Modern wireless communications require large sets of sequences with
good auto and cross-correlation. Preferably, such sequences should
be binary $(+/-1)$ or ternary $(+/-1,0)$ and have a high linear
complexity. Our matrices can be unfolded to yield such sets, and
many satisfy the above criteria. The unfolded sequences can be
utilized in CDMA. We consider two methods of matrix unfolding.
\subsubsection{Diagonal Unfolding}
Matrix $\textbf{A}=\{a_{i,j}\}$ of $T$ columns each of length $v$
with $gcd(v,T)=1$ can be unfolded along a diagonal
$s_{i}=\{a_{qi,ri}\}$ where $gcd(v,q)=1$ and $gcd(r,T)=1$. This is
because the diagonal passes through every entry in the matrix
exactly once, before repeating. Each one-dimensional cyclic shift of
the diagonal is equivalent to a two-dimensional $(k,l)$ cyclic shift
of the matrix. Therefore, correlation values for a matrix are equal
to those of the diagonal sequence. The above applies to exponential
and logarithmic shift sequences, because $gcd(p,(p-1)=1$. Such
matrices can be unfolded to yield new sequences of length $p^{2}-p$
directly along diagonals of such matrices [7], without changing the
correlation numbers.
\subsubsection{Row-by-Row Unfolding}
Any matrix $\textbf{A}=\{a_{i,j}\}$ of $T$ columns each of length
$v$ can be unfolded row-by-row into a sequence
$s_{i+jT}=\{a_{i,j}\}$ Here, there is no correspondence between
cyclic shifts of the long sequence and two-dimensional matrix
shifts. This kind of unfolding results in a doubling of the upper
bound on the number of matching columns [9]. The worst case
correlation are also doubled (approximately).
\section{Conclusion}
This paper shows how matrices with good two dimensional
autocorrelation and cross-correlation can be synthesized by using
cyclic shifts of pseudonoise columns. These sequences of shifts
(shift sequences) are derived from finite fields using mappings from
number theory. The properties of matrices constructed by shifts of a
pseudonoise sequence based on polynomial type shift sequences are
summarized in Table 1 below. Autocorrelation and cross-correlation
entries refer to numbers of matching columns. L refers to Legendre
sequence, H to Hall sequence and $M^{*}$ to M/GMW sequence of length
$2^{n}-1$ which is a prime number (Mersenne Prime). $M^{\dag}$
refers to non-binary M/GMW sequence.\\
\\
\small{
\begin{tabular}{|c|c|c|}
\hline
Construction $\varphi$ & Quadratic & Degree $n$ \\
\hline
Matrix Size $Z$ & $p\times p$  & $p\times p$\\
\hline
Total Matrices $N$ & $p-1$  &  $p^{n-1}-1$\\
\hline
Optimum Set $Q$ & $p-1$ & $p^{n-1}-1$ \\
\hline
Autocorrelation $\Theta_{AA}$ & $p,1,0$ & $p,n-1,n-2,...,0$ \\
\hline
Cross ($A,B \in Q )= \Theta_{AB}$ & $2,1,0$  & $n,n-1,...,0$ \\
\hline
Column Sequence $S$ & $L,H,M^{*}$ &$L,H,M^{*}$ \\
\hline
\end{tabular}}
\\
\begin{center} \textbf{Table 1 }\end{center}
Matrices based on exponential and inverse exponential (logarithmic)
shift sequences are summarized in Table 2. \small{
\begin{tabular}{|c|c|c|c|}
  \hline
   $\varphi$ & Exponential & Legendre & Zech \\
  \hline
  $Z$ & $(p-1)\times p$ & $p\times(p-1)$ & $(p^{m}-1)\times(p^{m}-1)$\\
    \hline
  $N$ & $ \phi (p-1) $ & $p-2$ &  $p^{m}-2$\\
    \hline
  $Q$& $2$ & $2$ & $2$ \\
    \hline
   $\Theta_{AA}$& $p-1,1,0$ & $p-1,1,0$ & $p^{m}-2,1,0$ \\
    \hline
  $\Theta_{AB}$& $2,1,0$ & $2,1,0$ & $2,1,0$ \\
    \hline
  $S$ & $L,H,M^{*}$ & $M^{\dag}$ & $M^{\dag}$ \\
  \hline
\end{tabular}}
\\
\begin{center} \textbf{Table 2 }\end{center}

\section*{Acknowledgment}
The authors thank the referees for many helpful suggestions,
including the re-kindling of our interest in window properties.

\end{document}